\documentclass[12pt]{article}
\usepackage{psfig}

\def\issue(#1,#2,#3){{\bf #1}, #2 (#3)} % AIP format

\def\opcit(#1){ {\em op. cit.}, #1}

\def\APP(#1,#2,#3){Acta Phys.\ Polon.\ \issue(#1,#2,#3)}
\def\ARNPS(#1,#2,#3){Ann.\ Rev.\ Nucl.\ Part.\ Sci.\ \issue(#1,#2,#3)}
\def\CPC(#1,#2,#3){Comp.\ Phys.\ Comm.\ \issue(#1,#2,#3)}
\def\CIP(#1,#2,#3){Comput.\ Phys.\ \issue(#1,#2,#3)}
\def\EPJC(#1,#2,#3){Eur.\ Phys.\ J.\ C\ \issue(#1,#2,#3)}
\def\EPJD(#1,#2,#3){Eur.\ Phys.\ J. Direct\ C\ \issue(#1,#2,#3)}
\def\IEEETNS(#1,#2,#3){IEEE Trans.\ Nucl.\ Sci.\ \issue(#1,#2,#3)}
\def\IJMP(#1,#2,#3){Int.\ J.\ Mod.\ Phys. \issue(#1,#2,#3)}
\def\MPL(#1,#2,#3){Mod.\ Phys.\ Lett.\ \issue(#1,#2,#3)}
\def\NP(#1,#2,#3){Nucl.\ Phys.\ \issue(#1,#2,#3)}
\def\NIM(#1,#2,#3){Nucl.\ Instrum.\ Meth.\ \issue(#1,#2,#3)}
\def\PL(#1,#2,#3){Phys.\ Lett.\ \issue(#1,#2,#3)}
\def\PRD(#1,#2,#3){Phys.\ Rev.\ D \issue(#1,#2,#3)}
\def\PRL(#1,#2,#3){Phys.\ Rev.\ Lett.\ \issue(#1,#2,#3)}
\def\SJNP(#1,#2,#3){Sov.\ J. Nucl.\ Phys.\ \issue(#1,#2,#3)}
\def\ZPC(#1,#2,#3){Zeit.\ Phys.\ C \issue(#1,#2,#3)}

\def\ET {\not\!\!{E_T}}
\def\be {\begin{equation}}
\def\ee {\end{equation}}
\def\bea {\begin{eqnarray}}
\def\eea {\end{eqnarray}}
\def\n {\nonumber}
\def\bc {\begin{center}}
\def\ec {\end{center}}
\def\t {\tilde}

\def \m3{\left |{m_{3}}\right |}

\def \H2{\left |{H_{2}}\right |}

\def \g2sq{{g_{2}}^{2}}

\def \MHD{m_{H_d}^2}
\def \MHU{m_{H_u}^2}
\def \MHF{m_{1/2}}
\def \MSX{m_{16}}
\def \MTN{m_{10}}

\def \MGL{m_{\tilde g}}

\def \MSL{m_{\tilde {\ell}_L}}
\def \MER{m_{\tilde {e}_R}}
\def \MTR{m_{\tilde {\tau}_R}}
\def \MTL{m_{\tilde {\tau}_L}}
\def \TL{\tilde {\tau}_L}
\def \LL{\tilde {l}_L}
\def \LR{\tilde {l}_R}
\def \MSR{m_{\tilde {\ell}_R}}
\def \SNU{\tilde\nu}
\def \MSNU{m_{\tilde \nu}}
\def \CH{\tilde \chi^{\pm}}
\def \MCH{m_{\tilde \chi^{\pm}}}
\def \S2F{sin^2{\theta}_{eff}}
\def\lapp{\mathrel{\rlap{\raise.5ex\hbox{$<$}}
                    {\lower.5ex\hbox{$\sim$}}}}
\def\gapp{\mathrel{\rlap{\raise.5ex\hbox{$>$}}
                    {\lower.5ex\hbox{$\sim$}}}}

\begin{document}
%\renewcommand{\thefootnote}{\fnsymbol{footnote}}
%%%%%%%%%%%%%%%%%%%%%%%%%%%%%%%%%%%%%%%%%%%%%%%%%%%%%%%%%%%%%%%%%%%%%
\title{
%\vspace*{0.10truein}
\begin{flushright}
\small hep-ph/0111222
\end{flushright}
%\vspace*{0.25truein}
Electroweak Precision Data, Light Sleptons and Stability of the SUSY Scalar Potential }
\author{ {\large\sl Amitava Datta}
\thanks{Electronic address: adatta@juphys.ernet.in}
\thanks{On leave of absence from Jadavpur University.}\\
Department of Physics, Visva-Bharati, Santiniketan - 731 235, India
\and
{\large\sl Abhijit Samanta}
\thanks{Electronic address: abhijit@juphys.ernet.in}\\
Department of Physics, Jadavpur University, Kolkata - 700 032, India}
\date{\today}

\maketitle

\begin{abstract}
The light slepton - sneutrino scenario with non - universal 
scalar masses at the GUT scale is preferred by the electroweak precision
data. Though a universal soft breaking mass at or below the Plank
scale  can produce the required non-universality at the GUT scale 
through running, 
such models are in conflict with the stability of the electroweak 
symmetry breaking vacuum. 
If the supergravity motivated idea of a 
common scalar mass at some high scale along with light sleptons is supported 
by future experiments that may indicate that we are living in a false vacuum.
In contrast SO(10) D - terms, which may arise if this GUT group breaks 
down directly to the Standard Model, can lead to this spectrum
with many striking phenomenological predictions,
without jeopardizing vacuum stability.
\end{abstract}

%\newpage
%%%%%%%%%%%%%%%%%%%%%%%%%%%%%%%%%%%%%%%%%%%%%%%%%%%%%%%%%%%%%%%%%%%%%%%%
%\pagestyle{plain}   \pagenumbering{arabic}
%\setcounter{footnote}{0}
\renewcommand{\thefootnote}{\arabic{footnote}}
%---------------------------------------------------------------------
%\section{Altarelli's Work and Motivation to Our Work}

The electroweak precision ( EWP ) tests by the experiments at 
LEP and SLC\cite{LEP} are on the whole in excellent agreement 
with the Glashow - Weinberg - Salam
standard model (SM) . However, if some judiciously chosen sub - set of the data is
examined, a few unsatisfactory features of the SM fit are revealed \cite{LEP,altarelli}

\begin{itemize}
\item The measured values of the parameter sin$^2\theta_{eff}$ from the observables
$A_{LR}$ and $A^b_{FB}$ differ at 3.5 $\sigma$ level.

\item Moreover, the  value of this  parameter as given by the hadronic asymmetries
and the leptonic asymmetries also exhibit a considerable discrepancy (at the
3.6 $\sigma$  level). 
  
\item When a global fit is performed a $\chi^2 / d.o.f $ = 26/15 corresponding to
a C.L. = 0.04 is obtained, which is hardly satisfactory.

\item  If the hadronic data is excluded from the global fit the quality
of the fit improves considerabily ( $\chi^2 / d.o.f $ = 2.5/3,  corresponding 
 C.L. = 0.48 ) while the exclusion of the leptonic data worsens the fit to an
unacceptable level ( $\chi^2 / d.o.f $ = 15.3/3 ,  corresponding   
 C.L. = 0.0016 ).
\end{itemize}

These observations tempt one to conclude  that the hadronic data may be plagued
by some hitherto unidentified experimental problem and, hence, the leptonic 
data should be taken more seriously \cite{altarelli}.

 This conclusion is challenged by the direct lower bound on the Higgs mass
$m_H > 113$ GeV\cite{higgs} and its indirect determination from EWP data
considering the leptonic asymmetries  only \cite{chanowitz,altarelli}. 
Using sin$^2\theta_{eff}$
measured from both hadronic and leptonic asymmetries, the  central value
of the fitted Higgs mass and the 95 \% C.L. upper limit on it happens to be 
98 GeV and 212 GeV respectively\cite{LEP}. These values consistent with the
direct search limit, have been confirmed by \cite{altarelli}. However,
if sin$^2\theta_{eff}$ from leptonic data only is employed, the corresponding
numbers become 42 GeV and 109 GeV, a situation which is hardly acceptable
vis - a - vis the direct limit. 
  
It must be admitted that there are uncertainties in the fitted value of
$m_H$\cite{altarelli}. The result has some sensitivity on the value of
$\alpha_{QED} (m_Z)$ which is scheme dependent although most of the  
existing schemes lead to upper bounds on $m_H$ in conflict with the direct search 
limit. Uncalculated higher order effects may have a  modest impact on  the fitted
value of $m_H$ \cite{altarelli}. Finally, if the current 1 $\sigma$ upperlimit
of the top mass ($m_t$ = 174.3 + 5.1 GeV ) rather than its central value is
used in the fit, then the compatibility of the fitted value of $m_H$ 
with the direct search limit improves.

Although  these uncertainties may conspire to produce an agreement
between the leptonic EWP data and the direct limit on $m_H$ within the 
framework of the SM, the situation is sufficiently provoking to 
reanalyse the data in extensions of the SM.

One  interesting possibility is to extend the discussion within the 
framework of supersymmetry \cite{susy}. Altarelli et al \cite{altarelli}
have found the MSSM parameter space ( PS ) where the SUSY  corrections 
to the elctroweak observables are sufficiently large and act in the 
direction of impoving the quality of fit.
The most significant loop contributions  come from the sneutrino $(\tilde \nu)$,
in particular if sneutrino mass is in the
in the range 55 - 80 GeV, and a perfect agreement with the data is obtained
with $m_H$ = 113 GeV. The charged left - slepton ($\LL$) mass is related
to $\MSNU$ by the SU(2) breaking D - term: ${\MSL^2 = \MSNU}_{\ell}^2 - 
\frac{1}{2}m_Z^2$cos2$\beta$, in a model independent way.   
Since it must be heavier than 96 GeV according to the LEP direct search limits
on charged sleptons \cite{aleph},
the parameter tan$\beta$ must be moderately large which is not a severe
restriction.

This spectrum, however, is incompatible with the popular
mSUGRA \cite{sugra} scenario with a common scalar mass $m_0$ at
the GUT scale $(M_G)$. Within the framework of mSUGRA such
light  sneutrinos  automatically demand even lighter right-sleptons, which
are already ruled out by the LEP mass  limits  on charged sleptons. Thus one has
to look for alternatives with nonuniversal scalar masses at $M_G$.
In this paper we shall look for such alternatives and scrutinize them 
in the light of vacuum stability.

We shall consider only those class of models where the sfermions of
the first two generations are nearly degenerate with mass  $m_0$ 
at $M_G$, as is required by the absence of flavour changing neutral currents. 
Moreover we shall assume a universal gaugino mass $\MHF$  at $M_G$
as this assumption is likely to be valid in all GUT models irrespective of the
specific choice of the GUT group. Given these parameters the left-slepton 
and sneutrino masses
of the first two generation at the weak scale can be computed by using the standard
one loop renormalisation group (RG)  equations. Other SUSY paramters may 
influence the running at the two loop level. Using ISAJET - ISASUSY 
we have convinced ourselves that these higher order corrections  are indeed negligible.
We constrain $m_0$ and $\MHF$ by requiring 55GeV $< \MSNU <$ 80 GeV at the 
weak scale (Fig. 1). 
The only other relevant SUSY parameter that enters the analysis
through the $SU(2)$ breaking D - term   is tan$\beta$,
although the dependence on it is rather weak.
Almost identical allowed PS is obtained for all tan$\beta \gapp$ 5.
As long as tan$\beta$ is not too large (say, tan$\beta  \lapp$ 20),
$\TL$ will be degenerate with the sleptons 
of the first two generations (to a very good approximation). For larger tan$\beta$, it may be
somewhat lighter. 
Since the experimental bound on the $\TL$ mass is considerably weaker 
($m_{\tilde {\tau}} >68$ GeV) than that for the selectron and smuon, 
higher values of tan$\beta$ can also 
be considered in principle, although we shall not pursue this case further.

The range of $m_0$ and $\MHF$ shown in Fig. 1 may be moderately 
altered if one considers a large hierarchy among the scalar masses 
at $M_G$. This happens due to the presence of a particular term in 
the RG equation which is usually neglected in the mSUGRA approximation 
(See eq. 4  and the discussions following it). We shall consider below 
a specific model with this feature.

So far no assumption about the other soft breaking parameters was 
necessary. However in order to take into account the chargino mass 
bound $\MCH >$ 100 GeV \cite{aleph} and to test the stability of the scalar 
potential\cite{oldufb,casas} 
one has to specify more SUSY parameters. In general $\MCH$ depends on
the Higgsino mass parameter $(\mu)$ and tan$\beta$ in addition to $\MHF$.  
The entire range of $\MHF$
in Fig 1 is such that $\mu$ can be chosen so as to make
$\MCH$ consistent with the LEP bound. Of course $\MCH$ is not a very 
sensative function of $\mu$ unless it is very small ($\mu \lapp$ 100 GeV). 
%{\bf CHECK}
We next turn our attention  
on $\MER$ and the stability of the scalar 
potential\cite{oldufb,casas}

%We extend the discussion of the data from MSSM \cite{altarelli}
%to SUGRA model in the SO(10) GUT. 

Before looking into specific models it is worthwhile to focus on some 
generic features of models with light sleptons.  
In several recent  works\cite{casas, paper1,paper2} on the  stability 
of the standard electroweak symmetry breaking ( EWSB ) vacuum, 
it has  been found that low mass
sleptons (to be more specific, sleptons significantly lighter than the 
electroweak gauginos) are
somewhat disfavoured. In view of the fact that there is already a strong 
lower bound on the chargino mass it is
important to check the compatibility  of the light sneutrino 
scenario favoured by the EWP data and vacuum stability.

The unbounded from below 3 ( UFB3 )  direction\cite{casas} of the scalar 
potential, its evaluation 
procedure and the choice of the generation indices $(i,j)$ which leads to the
strongest constraint are elaborately discussed in\cite{casas,paper1}.
To clarify why light sleptons are strongly disfavored, 
eqn. 93 of \cite{casas} has to be examined. The required equation is 
\be
V_{UFB3}=[\MHU+{\MSL}_i^2]\left |H_u \right |^2+\frac{\left|\mu\right|}
{\lambda_{e_j}}[{\MSL}_j^2
+{\MSR}_j^2+{\MSL}_i^2]\left |H_u\right |
-\frac{2{\MSL}_i^4}{g_1^2+\g2sq}.
     \label{ufbthree}
\ee
with $ i\neq j$. Here $\lambda_{e_j}$ is a leptonic Yukawa coupling and $g_1$ and
$g_2$ are the $U(1)_Y$ and $SU(2)$ gauge couplings respectively.
The UFB3 constraint arises from the requirement that  
$V_{UFB3}$ must be shallower than 
the EWSB minima $({V_0}_{min})$ (See eqn 92 of \cite{casas}). To get the strongest
constraints $i=1$ and $j=3$ is considered.
Over a large region of the PS corresponding to light sleptons,
the first term of eqn. 1 dominates  when
$\lambda_{\tau}$ is substituted in the second term.
%The minima of the UFB3 potential is usually obtained  for
%$H_2 \sim 10^6$GeV. 
The parameters are evaluated at a judiciously chosen
renormalisation scale $\hat Q$ where higher order loop 
corrections to the scalar potential are small and may be 
neglected\cite{gamberini,casas}. 
At this scale, the mass parameter $\MHU$ ($H_u$ refers to the Higgs 
bosons coupling to the up-type quarks) gets a large 
negative value which is required by radiative electroweak symmetry breaking 
(REWSB). Thus the fisrt term 
tends to violate the UFB3 constraint for small values of ${\MSL}_i^2$.
Infact it has been shown in reference\cite{paper2} that the 
anomaly mediated supersymmetry breaking ( AMSB ) model with light
sleptons violate the UFB3 constraint.

%The magnitude of the other terms are small by few orders of magnitude.
%This happens at 30 GeV less from the upper limit of common scalar mass
%$m_0$ at GUT scale. 
%Since only an order of magnidude estimate of $\hat Q$ is given in
%\cite{casas}, we have varied it from 0.1 $\hat Q$ to 10 $\hat Q$.
%This variation of  $\hat Q$  does not affect the conclusion.
%However, at the bondary  of allowed and disallowed region the second 
%term may become in competition.  In the case of PS allowed by
%UFB3, $\MHU$ gets negative value at a lower scale where second term
%plays the dominant role.
%It is found that squark and slepton masses get a stringent lower
%bounds, sometimes even  stronger than  experimental ones. 

We now wish to scrutinize the PS favoured by  EWP data ( Fig. 1 )
in the light of the stability of the vacuum. 
At this stage we have to be more specific about the model since 
$\MHU$, ${\MSR}_j^2$ and
$\left |\mu\right |$ are also needed to check this point. We first 
consider a SU(5) SUSY GUT with a common scalar mass $m_0$ at the Plank 
scale $( M_P \approx 2\times 10^{18}$ GeV )\cite{polansky} instead of $M_G$. 
An attractive feature of this model is that 
for the first two generations the mass of $\LR$ 
(denoted by $\MTN$ at $M_G$)
belonging to the $10$ plet of SU(5) happens to be larger than that of left
slepton belonging to the $\bar 5$ representation ( denoted by $m_5$ at $M_G$ )
due to the running between $M_P$ and $M_G$.
Thus the conflict between the low mass sneutrino and the LEP limit 
on $\tilde\ell_R$ mass seems to be resolved, at least qualitatively.

For the 3rd generation, $\MTN$ may be somewhat smaller if the 
relevant Yukawa couplings happen to be large at $M_G$ and contribute 
to the running (all relevant RG eqns are given in ref\cite{polansky}).  
This however, may not be a serious problem since
the limit on $\MTR$ is considerably weaker as dicussed above.

When we look into the numerical details the situation, however, is
far from simple.  According to Polansky et al the GUT scale values 
$\MTN$ and $m_5$ for the first two generations are 
approximately\cite{polansky} 
\be
\MTN^2 = m_0^2 + 0.45\MHF^2
\ee
\be
m_5^2 = m_0^2 + 0.30\MHF^2
\ee
 assuming that SUGRA generates the common scalar mass $m_0$ 
exactly at $M_P$. Since $\MHF$ has to be greater than 130 GeV
(approximately; see Fig. 1)  we find that $m_5$ is too large 
to give $\MSNU$ in the required range at the weak scale
even if $m_0 \approx 0$. We note that if the common
soft breaking mass is generated well below the Plank scale this 
difficulty may be avoided. Moreover GUT theshold corrections, which cannot be
computed precisely without specifying other GUT parameters like masses
of heavy multiplets,  may affect both $m_{10}$ and
$m_5$ to some extent. In view of these uncertainties  one can not discard
this model on this ground alone. We shall henceforth treat $\MTN$ and $m_5$
as  phenomenological parameters at $M_G$ with $\MTN > m_5$. Their actual values 
are to be chosen such that all charged slepton masses 
at the weak scale satisfy the LEP bound.

The Achilles' heel  of the model however, happens to be the running of $\MHU$
between $M_P$ and $M_G$. This running is controlled by not only the Yukawa couplings
$h_t$ and $h_b$ but also by $\lambda$ the coupling of the scalars belonging to the 
5, $\bar{5}$ and 24 plet of $SU(5)$. In course of running $\MHU$ is 
usually reduced as one goes below
$M_P$,  whereas $m_5$ driven by the gauge coupling alone increases. After considering 
various  scenarios
with different magnitudes of these  couplings ref\cite{polansky}  has concluded that 
$\MHU \lapp m_5$ in general,  while the equality holds if all the Yukawa couplings 
and $\lambda$ are negligibly small . We have checked that in such  scenarios
the UFB3 constraint is always violated for the PS in Fig. 1 
as is suggested by eqn. 1

Of course moderate shifts in $\MHU$ and $m_5$ may come from GUT threshold
corrections\cite{polansky} which may lead to $\MHU > m_{\bar 5}$. 
The magnitude of this shift depends on the details of the GUT model and we do not 
attempt to compute it. However,  adjusting $ m_5$ and 
$ \MTN$ such that both $\LL$  and $\LR$
satisfy the experimental bounds at the weak scale, we find that   
$m_{H_u},\MTN >>  m_5$ is needed to satisfy the UFB3 constraint 
(see the table for sample values). 
Such  large splittings between  $m_5$ and other GUT scale masses 
is unlikely to arise from theshold corrections.

If one considers an $SO(10)$ SUSY GUT instead, the matter fields of the 
first two generations
belonging to the 16 plet remain degenerate at $M_G$ even if running below $M_P$ is
considered. This will inevitably lead to a light $\LR$  at the weak scale if
the sneutrino mass is required to be in the range preferred by EWP data.

Thus running above the GUT scale alone in a SUGRA type scenario with  a common scalar
mass  generated between $M_P$ and $M_G$, is not likely  to yield the spectrum
preferred by EWP data if the stability of the vacuum is taken into account. 

If one gives up the UFB3 constraint by assuming 
that the standard vacuum is only a false vacuum \cite{falsevac},
while the global minimum of the scalar potential is indeed charge 
color breaking then the above constraints do not apply. 
If the tunnelling time for transition between the 
false vacuum and the true vacuum happens to be much larger than 
the age of the universe, such a model can not be rejected outright,
although it seems to be  against our intuitive notion of stability. 
Moreover the tunnelling  time, which can be routinely calculated in models with 
a single scalar, can not be computed reliably in models with multiple 
scalars. Yet the conclusions derived in the preceeding paragraphs do not
loose their significance. If future experimental data confirms light 
sleptons along with a mass spectrum stemming from a SUGRA motivated 
common scalar mass at some high scale 
$\lapp M_P$, then that would indicate that we may be living in a false vacuum, 
no matter how counter intuitive it may appear to be at the first sight.

The remaining of this paper shall deal with a type of non-universality
which arises when a GUT group breaks down to a group of lower rank leading
to non-universal D-terms at $M_G$ \cite{d-term}. This type of models 
can produce the spectrum preferred by EWP data without violating the
UFB3 constraint. As a specific example we
consider an $SO(10)$ SUSY GUT  breaking down to the SM in a single step.
The relevant mass formulae at $M_G$ are:
\bc
$m_{\tilde Q}^2 = m_{\tilde E}^2 = m_{\tilde U}^2 = m_{16}^2 +
m_D^2$\\ $m_{\tilde D}^2 =m_{\tilde L}^2 =m_{16}^2 -3 m_D^2$\\
$m_{H_{d,u}}^2 =m_{10}^2 \pm 2 m_D^2$\\
\ec
where $m_D$ is the D-term with unknown magnitude,  
the common mass of all the members of the 16-plet of $SO(10)$ 
at $M_G$ is denoted  by $\MSX$ and the common Higgs mass  by $\MTN$.

This model is interesting since even though  all  sfermion masses 
are degenerate at $M_G$,which indeed should be the case for the first two
generations of sfermions as discussed above, the D - terms  may  introduce significant 
nonuniversality between the L and R - sleptons making the latter
somewhat heavier than the former. 
Thus a light sneutrino as required by the EWP 
data does not necessarily imply a lighter R- slepton. 

In general the Higgs mass $\MTN$ and $\MSX$ could be different at $M_G$
due to the running between $M_P$ and $M_G$. However, it is interesting to
note that even if $\MTN$ and $\MSX$ are nearly degenerate at $M_G$, the
D-term may make $\MHU$ significantly heavier that the left sleptons at $M_G$.
Because of this reason the model can be UFB3 stable without requiring $\MTN$ to
be much larger than $\MSX$.
%%%%%%%%%%%%%%%%%%%%%%%%%%%%%%%%%%%%%%%%%%%%%%%%%%%%%%%%%%%%%%
%\section{OUR RESULTS}
We shall consider both universal ($\MSX = \MTN$) 
and nonuniversal ($\MSX \ne \MTN$) scenarios.
%For this Altarelli et al found $\MSNU$'s in the range 55 - 80 GeV.

The methodology of finding the spectra is same as in \cite{paper1}.
$\mu$ and $B$ are determined by the REWSB condition at a scale 
$M_S = \sqrt{m_{{\tilde t}_L}
m_{{\tilde t}_R}}$. Then we put the experimental constraints. 
For a given $\MSX$ and $\MHF$, we consider the smallest $m_D$ such that 
$\MSNU <$ 80 GeV. Larger values of $m_D$ may also be considered 
provided $\MSNU$ is 
in the range 55 GeV $< \MSNU<$ 80 GeV.
However, larger values of $m_D$ tends to yield  stronger UFB3 constraints. 

We first discuss  the APS  without requiring Yukawa unification,
in the $\MSX - \MHF$ plane for $\MSX = \MTN$,
$A_0 = 0$, tan$\beta$ = 15 and $\mu >$ 0
as shown in Fig. 2. The upper bound on $\MHF$ for a given $\MSX$ corresponds to the 
situation when no $m_D$ can give ${\MSNU}_{e,\mu} \le 80$ GeV
and the lower bound by experimental lower limit on  $\CH$.
The D-term can control $\MSL$ and, hence $\MSNU$, over a large range
of $\MSX$, which, therefore, is found to be large.
If we increase $\MSX$ further, the contribution from $\tau$ Yukawa coupling
 decreases $\MTL$ even for tan$\beta$ = 15 thanks to a large $\MSR$. 
As a result ${\MSNU}_{\tau}$ falls below the 
experimental bound (43.6 GeV), even though  ${\MSNU}_{e,\mu}$ are 
in the viccinity of  80 GeV. 
The  upper and lower limits  on $\MSX$ significantly
depends on $A_0$ and tan$\beta$.
 
The fact that the allowed range of $\MHF$ increases with $\MSX$ is rather
puzzling. The origin of this lies in a term in the RG eqn which is ususally
neglected in mSUGRA.

%F(16)=FAC*(-.6*G(1)**2*G(7)**2-3*G(2)**2*G(8)**2-.3*G(1)**2*SMV)
%SMV=G(14)-G(13)+(2*G(19)+G(24))-(2*G(16)+G(21))
%         -2*(2*G(18)+G(23))+(2*G(17)+G(22))+(2*G(15)+G(20))

\bea
\frac{d\MSL}{dQ}=\frac{3}{8\pi^2} [- 0.6 g_1^2 M_1^2 - 3 g_2^2 M_2^2~~~~~~~~~
~~~~~\n\\ 
- 0.3 g_1^2
\{\MHU - \MHD + (2 m_{\t u_L}^2 + m_{\t t_L}^2)\n\\
- (2 m_{\t e_L}^2 + m_{\t\tau_L}^2)
-2 (2 m_{\t u_R}^2 + m_{\t t_R}^2)\n\\
+ (2 m_{\t d_R} + m_{\t b_R})
+ ( 2m_{\t e_R} + m_{\t\tau_R}^2)\}]
\eea

The last term on the right hand side is zero at $M_G$ in the mSUGRA model. Moreover its
coefficient is rather small. Hence the contribution of this term 
remains small even at the weak scale. In the D - term model, however, 
this term is already large at the GUT scale
in particular due to the $\MHU - \MHD$ term. This difference is indeed large if the
D-term is chosen to be large in order to have $\MSNU$ in the desired range. The slepton
and sneutrino masses are  reduced under the influence of this term by as much at 10 - 15 GeV
for large $\MSX$. As a result unexpectedly large values of $\MHF$ can be accommodated. 

If tan$\beta$ is lowered, the mass of lightest Higgs($m_h$) 
decreases rapidly, low values of $\MSX$ are not allowed if $m_h \gapp$ 113 GeV
is required. 
However, if $\MSX$ is increased, the higgs mass increases appreciably
through  radiative corrections. Moreover the running of $\MTL$ and hence of
$m_{\t\tau_\nu}$, are also modest for low tan$\beta$. Due to these reasons  
 higher values of $\MSX$  are allowed. We find 300(60) GeV $\lapp\MSX 
\lapp$ 700(460) GeV for tan$\beta$=7(15), while the  other parameters are the same as 
in Fig. 2.

Increasing 
the absolute value of $A_0$ makes large 
difference  between ${\MSNU}_{e,\mu}$ and ${\MSNU}_{\tau}$.
As a result $\MSX$ gets a stringent upper bound. 
It also lowers $m_H$
very rapidly giving a strong lower bound on $\MSX$.
For  example,
$60(120)\lapp \MSX \lapp 460(420) GeV$ for $A_0 =0(\MSX)$, the 
other parameters being the  same as in Fig. 2. 

There are also appreciable  changes  in the APS with change in the sign of $\mu$. 
The masses $\MCH$ and $\MTL$ increase significantly  as one change $\mu < 0$ to
$\mu > 0$. One need high value of $\MHF$ to keep $\MCH$ above experimental
bound and high value of $\MSX$ for $\MTL$ above experimental 
bound for $\mu < 0$. For example $60(140)\lapp \MSX \lapp 460(440) GeV$ 
for $\mu >0(<0)$, while the other parameters are as in Fig.2.

We next examine the UFB3 constraint for the APS in Fig. 2. One of the 
important conclusions  of this paper is that
the UFB-3 constraint rules out the entire  APS
for the universal model (throughout this paper we shall use a $\ast$ (+) 
for a UFB3 disallwed (allowed) points in the PS).

Next  we will consider the effect of nonuniversality (compare Fig. 2 
and Fig. 3). The SUSY parameters in Fig 3 are as in Fig 2 except that
$\MTN$ = 1.5 $\MSX$. Such a modest non - universality may arguably appear
due to threshold corrections at $M_G$. 
For higher values of $\MTN$ 
 $\mu^2$ decreases rapidly and $\MCH$ comes below experimental
bound. A larger  $\MHF$ can avoid this problem  but then the constraint 
$\MSNU <$  80 GeV requires a D - term that makes sfermion mass square negative
at GUT scale. The overall APS, therefore, decreases. 
 However, a region is still UFB3 allowed  for $A_0 \gapp 0$, since $\MHU$ 
is somewhst larger at $M_G$ to begin with.

Next we consider the possibility of Yukawa unification 
in this model\cite{partial}. It has already been shown in \cite{paper1,tata}
that full $t - b - \tau$ Yukawa unification does not permit low
slepton masses even in the presence of D-terms. We shall, therefore, 
restrict ourselves to  partial $b -\tau$ unification with an accuracy $\le 5\%$.
We fix  tan$\beta$ to its lowest value required by unification.
The APS in the universal model (Fig 4) is qualitatively the  same 
as in the fixed tan$\beta$ case 
(compare Fig. 2 \& Fig. 4)
but its size somewhat smaller. It has been 
found that for higher values of $m_D$ unification requires relatively
low values  of tan$\beta \sim 20 $ . As indicated  in Fig 4  
the APS is not consistent with the UFB3 constraint. 
Introduction of a modest non-universality at $M_G$ as before, reduces the APS
but leads to several  UFB3 allowed points ( Fig. 5).
The following observations in the context of this model are worthnoting.
%\begin{itemize}
%\item
%We find that univerasal case ($\MTN = \MSX$) is totally ruled 
%out by the stability criterion of the scalar potential.
%\item
%A small APS  allowed by UFB is obtained for $A_0 > 0$ and $\MTN \gapp
%1.5\MSX$. 
%\item
% the squark, $\MSR$ masses can be made as high as 1 TeV.
%\item
% The light $\MCH$ and $\MSL$ $m_{\tilde \tau}$ are as low as 100 GeV.
%\item
i) We find a strong lower bounds $\MER \gapp $225 GeV and 
$m_{d_R}\gapp$ 320 GeV  from the UFB3 constraint.
ii)
We get a tight upper bound of tan$\beta \lapp$30 independent of the 
choice of other
parameters.
%\item
%We find from our earlier work\cite{paper2} in mAMSB model the APS which 
%is allowed by  EWP data is completely ruled out by UFB3 constraints.
%\end{itemize}
%In conclusion we demand that this $SO(10)$ D-term model is very 
%promising model to test supersymmetry in the present colliders.
%SIGNALS ?

%Discussions on heavy higgs mass

%WILL WE DISCUSS AMSB MODEL IN THIS PAPER ?

The phenomenological significance of a light sneutrino has already been discussed at 
length  in the literature[18-24]
%admanobiswarup,admanosuro,adsreerup,
%admanonirmalya,akdmanonirmalya,adshyama,adparida,addrees, addreesmano}. 
If the sneutrino mass happens to be in the range
preferred by EWP data then it decays into the invisible channel 
$\SNU \rightarrow \nu{\chi_1}^0$ with 100\% BR and becomes an additional carrier
of missing energy. If the lighter chargino mass happens to be near the current 
lower bound, a situation also preferred by EW precision data, then it would 
decay into the channel $\CH  \rightarrow \ell\SNU$ with almost 100\% BR 
(the decay into sleptons are phase space suppressed), while 
in the coventional mSUGRA scenario it dominantly decay into jets.
Finally the second lightest neutralino ${\t\chi_2}^0$ which happens to be nearly
degenerate with $\CH$ in models with gaugino mass unification, 
also decays dominantly
into the invisible channel ${\t\chi_2}^0 \rightarrow {\t\chi_1}^0\SNU$ and becomes
another source of missing energy. 

The additional carriers of missing energy  which play roles similar to that of the
lightest supersymmetric particle ( LSP ), may be termed virtual LSP(VLSP) in the context of
collider experiments \cite{admanobiswarup}

In the VLSP scenario the collider signatures of squark - gluon production are quite 
different from the ones in conventional mSUGRA model due to the 
additional carriers of 
missing energy. Moreover thanks to the enhanced leptonic decay of the chargino the
lepton + jets + $\ET$ signal may increase at the cost of  jets + $\ET$
signature \cite{admanobiswarup,admanonirmalya} The hadronically quiet 
tri-lepton signature
\cite{admanobiswarup} signalling the $\CH {\t\chi_2}^0$ production 
at the hadron colliders 
may disappear due to the invissible decay of ${\t\chi_2}^0$. On the other 
hand the hadronically quiet dilepton + $\ET$ signal from 
$\tilde\chi^\pm \tilde\chi^\mp$  may be boosted at the upgraded
Tevatron as well at the $e^+e^-$ colliders due to the enhanced leptonic 
decays of charginos\cite{admanosuro,addreesmano}. Another dramatic 
signal of the VLSP model could be increase in the $e^+e^- \rightarrow \gamma$ +
missing energy events\cite{adsreerup}. In the conventional mSUGRA model the SUSY contributions
comes only from the channel $e^+e^- \rightarrow 
\nu {\t\chi_1}^0{\t\chi_1}^0$ which has a modest cross section and is often swamped by
the  $e^+e^- \rightarrow \gamma \nu \bar{\nu}$ background. 
In the VLSP scenario, however, $e^+e^- \rightarrow 
\gamma \t{\nu} \t{\nu}^*, \gamma {\t\chi_1}^0{\t\chi_2}^0, 
\gamma {\t\chi_2}^0{\t\chi_2}^0$
contributes to the signal in addition to the above conventional mSUGRA process.
Implementing some special cuts devised in \cite{adsreerup} one can easily suppress
the SM background. In particular a suitable cut on the photon energy may kill a
large number of $e^+e^- \rightarrow \gamma \nu \bar{\nu}$ events arising due to
the radiative return to the Z peak at  LEP energies above the Z pole
without affecting the signal. A reanalysis 
of the LEP data using such cuts may  reveal the VLSP scenario or severly
restrict the sneutrino mass range preferred by EWP data.

If $m_{t_1} < \MCH$, then the preferred decay mode of the lighter stop ( $\t t_1$ )
could be $\t t \rightarrow b \ell \t \nu$ rather the loop induced decay 
${\t t} \rightarrow c {\t\chi_1}^0$ \cite{admanonirmalya}. This would enhance the leptonic signal
from the stop at the cost of jets + $\ET$ events.

While light sleptons may arise in many scenarios including the ones 
not based on supergravity (e.g., in the AMSB model), the simultaneous 
presence of relatively right down squarks and light sleptons would 
vindicate the SO(10) D-term model. Enhancement  of the jets + missing energy 
signal at the expenses of leptons + jets + $\ET$ signal from squark
gluino production would be the hall-mark of this scenario\cite{adparida,addrees,
adshyama}. The effect becomes particularly striking if $\MGL > m_{\tilde d_R}$,
while all other squarks are much heavier than the gluinos\cite{addrees,adshyama}.
This mass hierarchy is infact obtained over the bulk of the parameter space 
probed in this paper.

\begin{table}[htbp]
\caption{\sl{ Sample GUT scale masses and consistency with the 
UFB3 condition for  $A_0$ = 0, tan$\beta$ = 15, $\mu >$0, $\MHF = 153$ GeV.
}}
%\vspace*{2mm}
%\hskip4pc\vbox{\columnwidth=26pc
\begin{center}
\begin{tabular}{|c|c|c|c|c|c|}
\hline
$m_{\t e_L}$ & $m_{\t e_R}$ & $m_{H_u}$ & $m_{H_d}$ & \\
(GeV) & (GeV) & (GeV) & (GeV)&  \\
\hline
36       &  210 & 240  & 340  & UFB3 allowed  \\
36       &  210    & 220  & 340     & UFB3 disallowed\\
 36       &  210   &  240    &  300  & "    \\
           
\hline
\end{tabular}
\end{center}
%}
\end{table}
%-------------

{\em \bf Acknowledgements}:

The work of AD was supported by DST, India (Project No.\ SP/S2/k01/97).
AS thanks CSIR, India, for his research fellowship.

%\newpage

%---------------------------------------------------------------------
\begin{figure}[htb]
\centerline{
\psfig{file=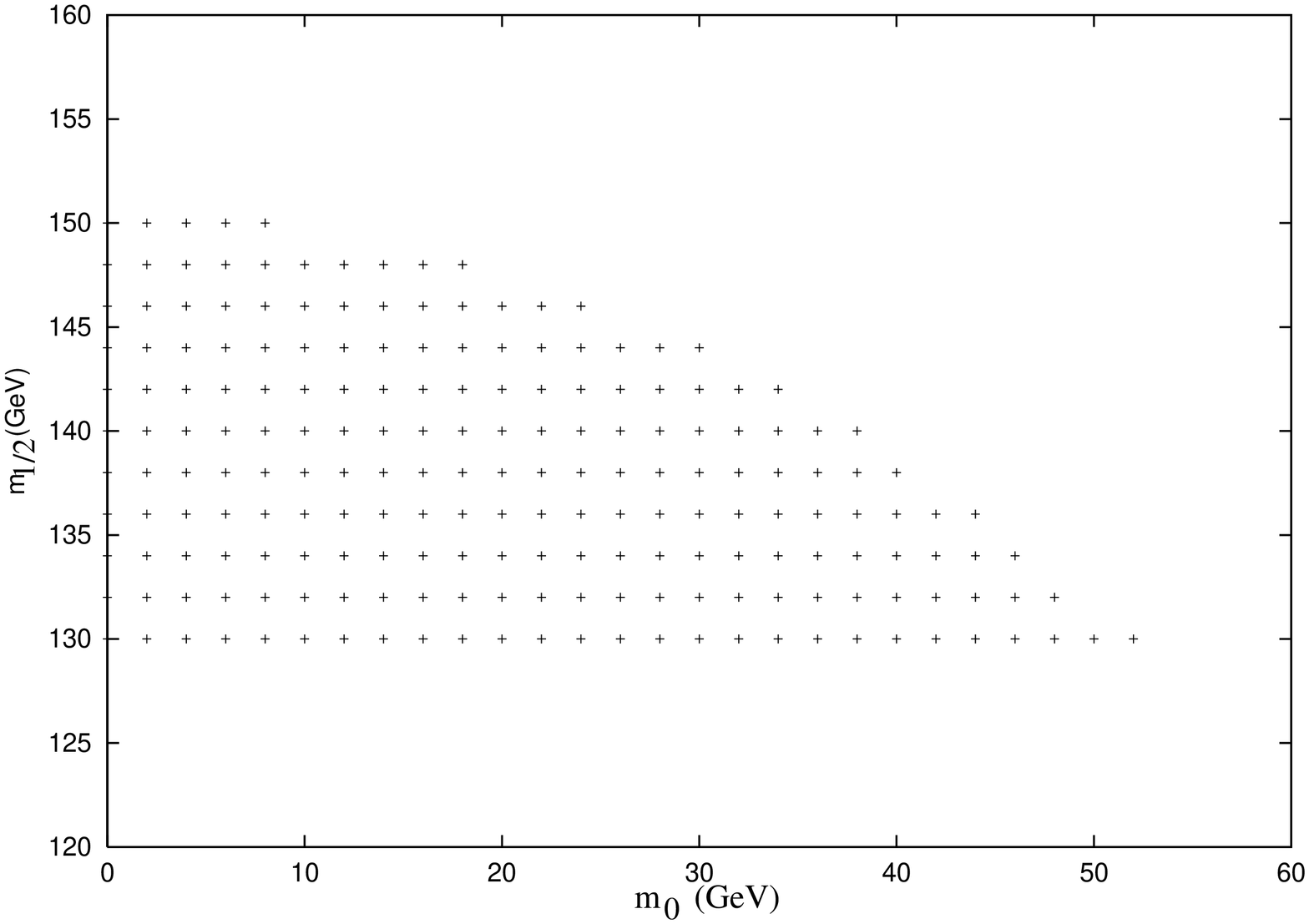,width=12cm,angle=270}
}
\caption{\sl{The APS in the $m_0-\MHF$ plane for 55GeV$<\MSNU<$80GeV with 
tan$\beta$ = 15. The 
lower limit on $\MHF$ is due to the
chargino mass bound from LEP.        }}
     \label{Fig. Fig7}
\end{figure}
\vspace*{5mm}
\begin{figure}[htb]
\centerline{
\psfig{file=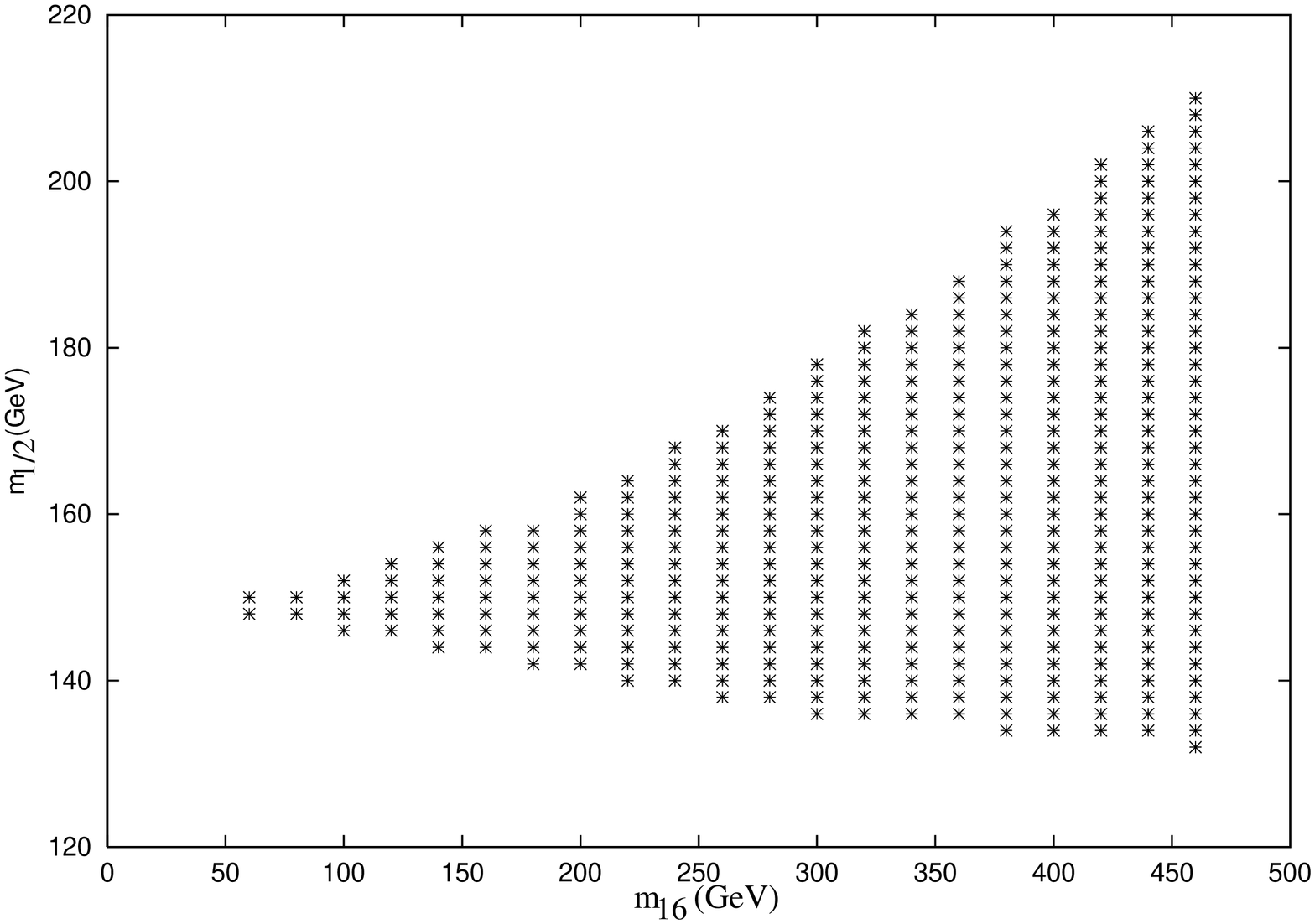,width=12cm,angle=270}
}
\caption{\sl{The APS for 55GeV$<\MSNU<$80GeV in the SO(10) model with
 $\MTN = \MSX,  A_0 = 0$, tan$\beta$ = 15 and 
$sign(\mu)>0$ and $m_D$ is fixed by the light sleptons 
criterion. In our notation a $\ast$ denotes a point ruled out by UFB3 while
a + indicates a UFB3 allowed point.
           }}
     \label{Fig.figure7}
\vspace*{5mm}
\centerline{
\psfig{file=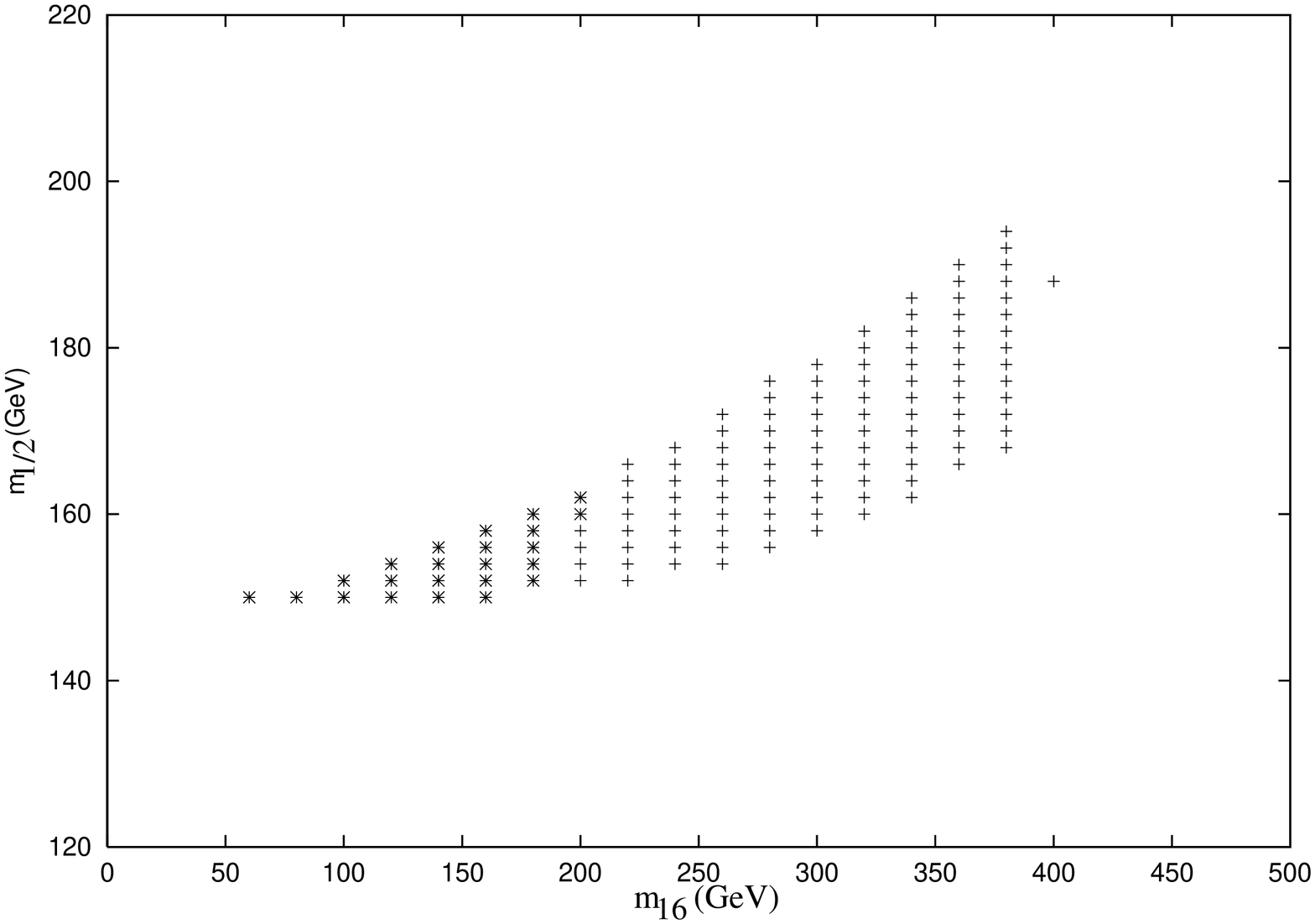,width=12cm,angle=270}
}
\caption{\sl{The same as Fig. 1, with $\MTN = 1.5\MSX$.
           }}
     \label{Fig.fig8}
\end{figure}
\begin{figure}
\vspace*{5mm}
\centerline{
\psfig{file=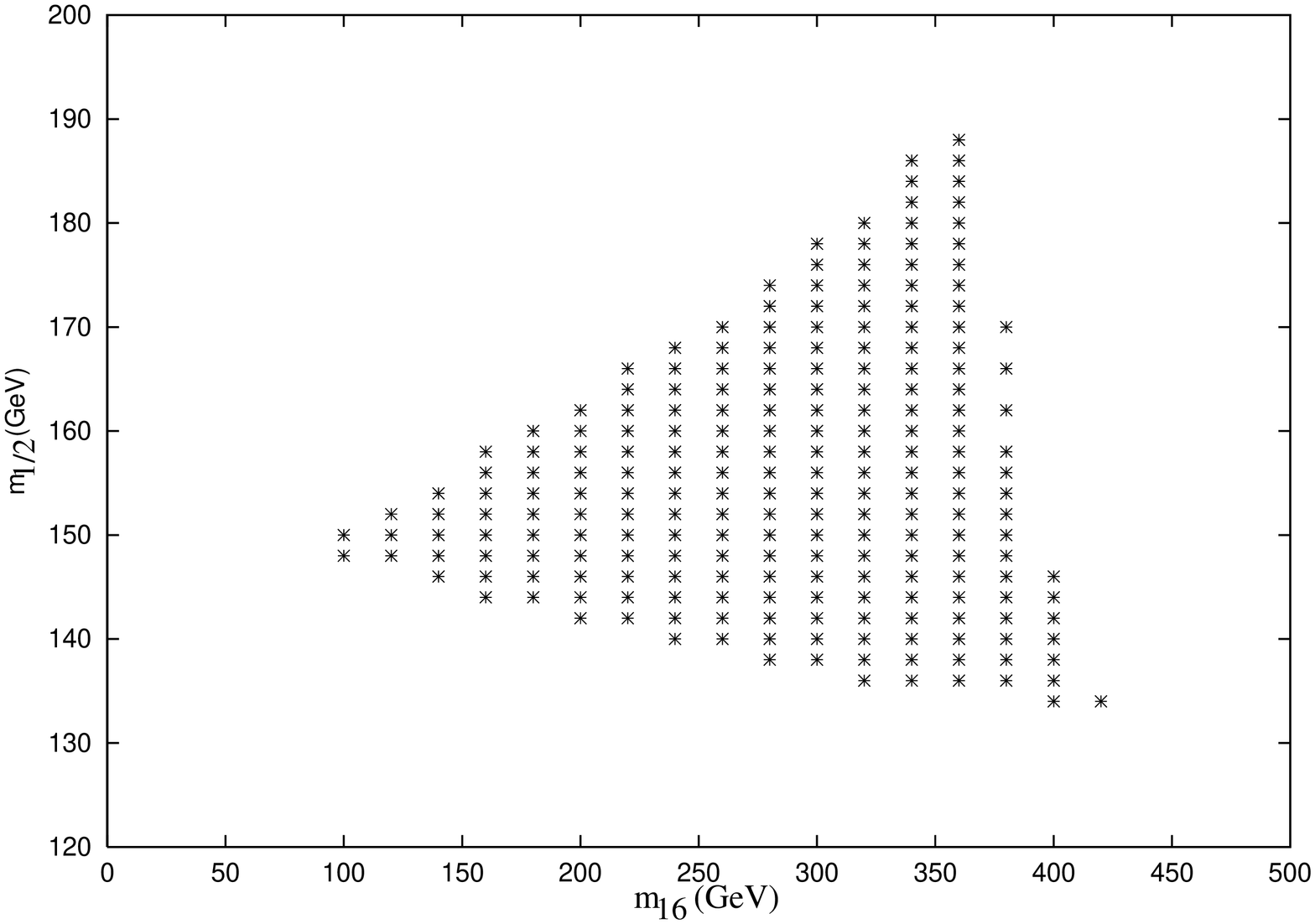,width=12cm,angle=270}
}
\caption{\sl{The allowed parameter space in the universal scenario with
             $b-\tau$ unification. 
             We set $A_0=0$ and $m_D$ is fixed by the light slepton 
             criterion.
              All points allowed
             by the Yukawa unification criterion are 
             ruled out by UFB3.
            }}
     \label{Fig.fig8}
\vspace*{5mm}
\centerline{
\psfig{file=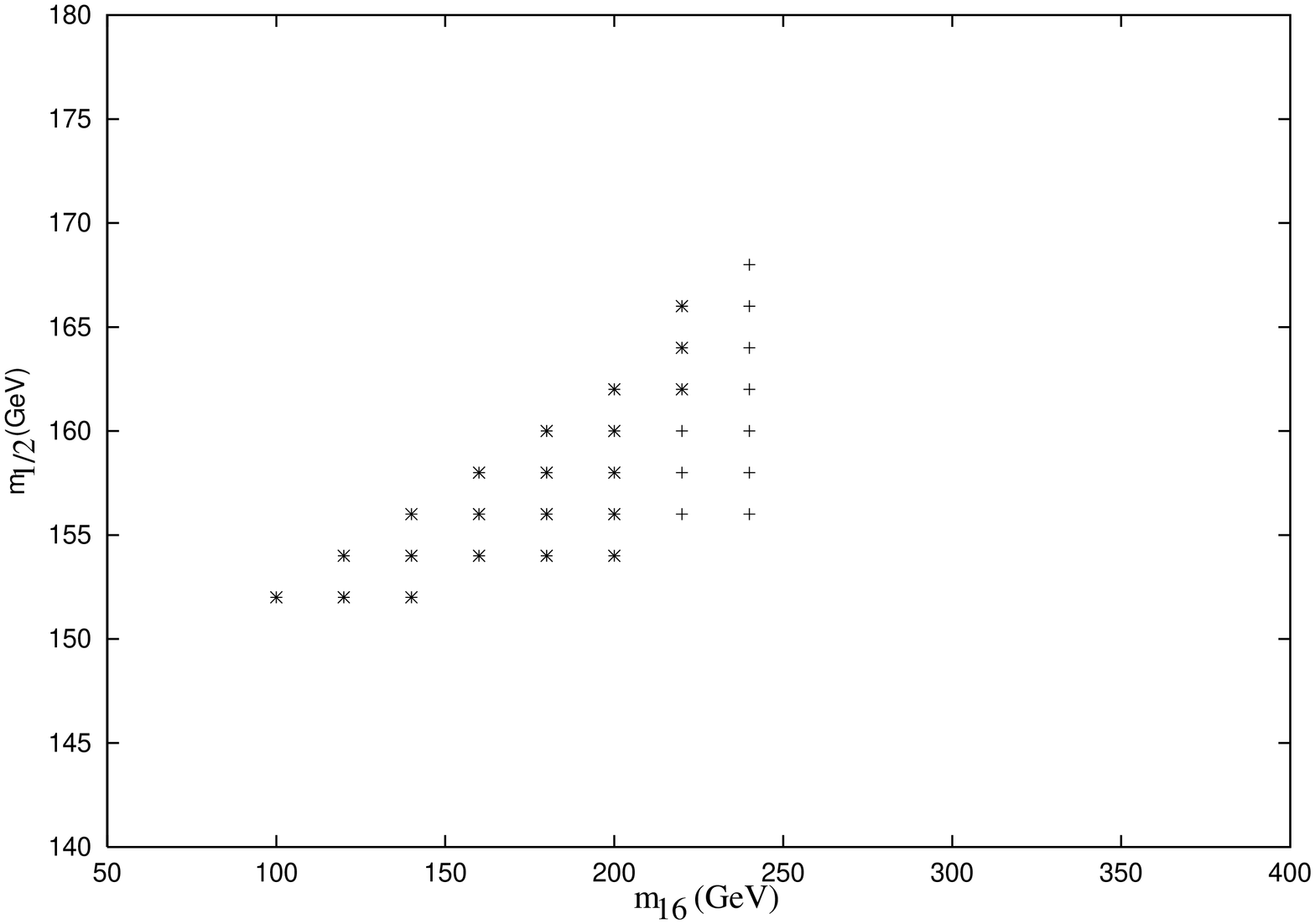,width=12cm,angle=270}
}
\caption{\sl{The same as Fig. 3, with $\MTN = 1.5\MSX$.
           }}
     \label{Fig.fig8}

\end{figure}
%
%\newpage

\end{document}